 \documentclass[final,3p,times,twocolumn]{elsarticle}

\usepackage{graphicx}

\usepackage{amssymb}

\usepackage{lineno}

\journal{Nuclear Instruments and Methods A}

\begin{document}

\begin{frontmatter}



\title{Acoustic detection of ultra-high energetic neutrinos
\linebreak
 - a snap shot - }


\author{Rolf Nahnhauer}

\address{DESY, Platanenallee 6, D-15738 Zeuthen, Germany}

\begin{abstract}

Already more than 30 years ago the acoustic particle detection method has been
 considered to be one possibility to measure signals from ultra-high energetic
 neutrinos. 
The present status and problems of corresponding model predictions are
 discussed  in comparison with existing experimental measurements. 
Available acoustic sensors and transmitters are described and new ideas for
 corresponding applications are mentioned. 
Different methods for in-situ calibrations are discussed. 
Results of measurements of acoustic test arrays at different sites are
 presented in some detail. Future activities for applications of the
 technology in large size detectors are evaluated.

\end{abstract}


\end{frontmatter}

\linenumbers

\section{Introduction}

The detection and study of ultra-high energetic neutrinos above $10^{17}$ eV
 created in cosmic sources or by interaction or decay of even higher energetic
 particles became of increasing interest during the last decade. 
Several interesting questions of particle physics, astrophysics and cosmology,
 could be studied measuring interactions of such neutrinos on Earth with
 reasonable statistics \cite{UHE-Phys}.

Corresponding neutrino flux predictions were already small ten years ago
 \cite{ESS} but had to be decreased further by new bounds from cosmic ray
 \cite{Auger-1} and  high energy gamma ray measurements \cite{Fermi-1}. 
Present day experiments could derive therefore until now only flux limits
 \cite{IceCube-1,Auger-2,RICE,ANITA}. 
A first detection of such neutrinos is expected with detectors of $\sim$100
 ${\rm km^3}$ size. 
To get reasonable statistics will need probably about an order of magnitude
 larger detector volumes. 
It seems impossible today to instrument such experiments with conventional
 optical detectors within reasonable cost limits. 
Acoustic particle detection may be one option among others to overcome this
 problem \cite{ARENAS}. 

The possibility to detect charged particles by the sound they produce passing
 through matter was the first time mentioned in 1957 by G. Askaryan
 \cite{ASK-57}. 
About 20 years later a corresponding model was formulated
 \cite{ASK-76,BOW-76}, and first ideas about a 100 ${\rm km^3}$ acoustic
 detector were discussed seriously \cite{AD-77}. 
In the following sections it will be shown, how far the predictions of the
 Thermo-acoustic Model could be confirmed experimentally and what questions
 have still to be answered.

During the last 10 years different groups tried to use acoustic test arrays to
 extract basic information needed to build large scale detectors in different
 materials and environmental conditions. 
Their results will be summarized in the second part of this paper. 
Finally future steps for the improvement of the acoustic technology and its
 application will be mentioned.  

\section{The Thermo-Acoustic Model}

Ideas about the Thermo-acoustic Model of the creation and propagation of sound
 in particle interactions were displayed for the first time at the 1976 DUMAND
 meeting \cite{DUM}. 
They were formulated independently by Bowen \cite{BOW-76} and Askaryan and
 Dolgoshein \cite{ASK-76}. 
More detailed descriptions of both concepts were published three years later
 \cite{ASK-79,LEA-79}. 
Recently a new approach based on \cite{LEA-79} has been published including
 signal attenuation effects \cite{ACO-1}.

In neutrino interactions a charged or neutral lepton and a hadronic particle
 cascade is produced. 
The cascade gives rise to a large energy deposition in a small volume in a very
 short time. 
The volume is overheated what gives rise to a pressure wave which developes
 orthogonal to the cascade and therefore the incident neutrino direction.

 \subsection{model predictions}

An illustrative way to describe the dependencies of the important quantities
 and variables of the process is given in \cite{ASK-79}.
\begin{eqnarray}\label{eq1}
  p &=& (k/c_p) (E/R) M \nonumber\\
  M &=& (f^2/2)(\sin{x}/x)\nonumber\\
  f &=& v_s/(2d) \nonumber\\
  x &=& (\pi L/2d)(\sin{\delta})
\end{eqnarray}
with $p$ : pressure amplitude, $E$ : cascade energy, $R$ : distance to receiver,
 $f$ : frequency, $v_s$ : speed of sound, $d$ : cascade diameter,
 $k$ : volume expansion coefficient, $c_p$: specific heat, $L$ : cascade length,
 $\delta$ : angle between normal to cascade direction and receiver.

The model predicts a linear dependence of the pressure amplitude on the
 particle cascade energy, which is related to the incoming neutrino energy and
 allows to determine the neutrino direction from measuring the pressure wave
 propagation through the medium.

\begin{figure}
  \centering
  \includegraphics[width=8.0cm]{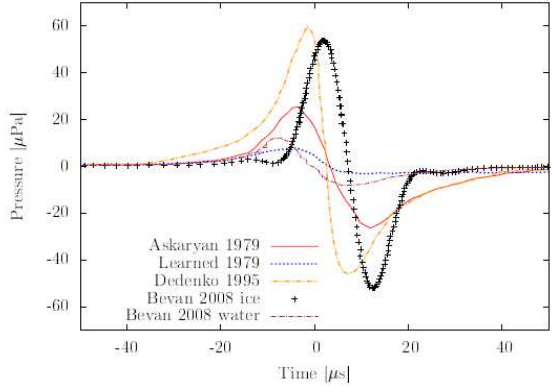}
  \caption{\label{fig:signal-butkevich} Signal strength from three
    different acoustic model parameterizations \cite{BUT} and inclusion of attenuation effects \cite{ACO-1} (from \cite{DESC})}

\end{figure}

Unfortunately absolute signal predictions are uncertain mainly due to the not
 well known particle interaction and energy loss process at ultra-high
 energies. 
Different assumptions about what values should be used for the width and the
 length of a cascade of a certain energy give rise to signal predictions which
 differ by about an order of magnitude (see  fig.~\ref{fig:signal-butkevich}).

In the considered energy range two other phenomena have also to be taken into
 account: the Landau-Pomeranchuk-Migdal-effect and photo- and electro-nuclear
 reactions \cite{KLEIN}. 
The LPM-effect predicts decreasing cross section for bremsstrahlung and pair
 production, elongating hadronic cascades above about $10^{18}$ eV whereas the
 second effect act in the opposite direction at energies above $10^{20}$ eV.  

\subsection{target material effects}

One could rewrite eq. \ref{eq1} in the form
\begin{equation}
  p = \gamma ~(E/R)~ M'
\end{equation}
with $\gamma = v_s^2 ~(k/c_p)$ and $M' = (1/2)(1/d^2)(\sin{x}/x)$.\\
The Grueneisen coefficient $\gamma$ is a strongly material dependent quantity. 
In tab. \ref{fig:tab-1} this is displayed for three materials under discussion
 for acoustic detector applications. 
At the same incoming energy signals in ice should be therefore nearly an order
 of magnitude larger than those in  water. 
In salt even larger signals are expected. 
Also permafrost was recently suggested to give rise to quite large signals
 \cite{NRT}.

\begin{table}[hbtp]
  \caption{\small Thermo-acoustic model parameters and boundary conditions for
    three different materials (adapted from \cite{Price})
     \label{fig:tab-1}}
   ~\\[-2mm]{      
     \begin{tabular}{|l|c|c|c|} \hline
       & water & South Pole ice & salt \\ \hline \hline
       $c$ (m/s) & 1530 & 3880 & 4560 \\ \hline
       $(k/10^{-5}) [{\rm K^{-1}}]$ & 25.5 & 12.5 &
       11.6 \\ \hline
       $c_p [{\rm J (K~kg)^{-1}}]$ & 3900 & 1720 & 839 \\ \hline
       $\gamma$ & 0.153 & 1.12 & 2.87 \\ \hline
       $f_{max} {\rm [kHz]}$ & 7.7 & 20 & 42 \\ \hline
       refraction & moderate & very small & small? \\ \hline
       $\lambda_{\rm att}$ & $>$1000 m & $\sim$300 m & $>$100 m \\ \hline
       noise & variable & stable,$<$14 mPa & small? \\ \hline
     \end{tabular}
    }
 \end{table}

For water things are even more complicated, because the volume expansion
 coefficient for water depends strongly on the water temperature. 
At 4 degree Celsius it is equal to zero having different signs below and above
 this temperature. 
The signal strength for acoustic pulses in water depends therefore strongly on
 the specific location of the neutrino interaction and may vary also by a
 factor 10 (see \cite{BUD,SUL}). 

\section{Experimental verifications}

\begin{figure}
  \centering
  \includegraphics[width=8.0cm]{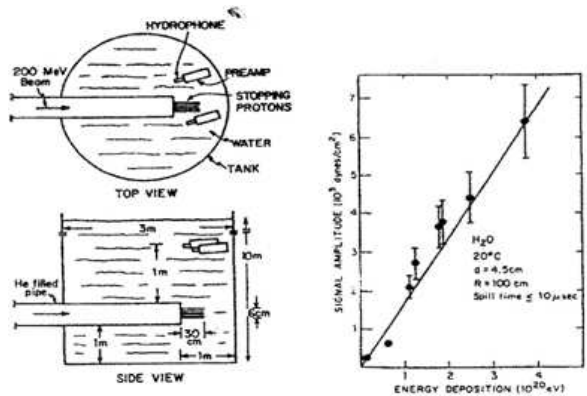}
  \caption{\label{fig:sulak} Experimental setup and results for signal dependence on deposited energy for an early Thermo-acoustic model check (from \cite{SUL})}

\end{figure}

Already at the end of the seventies of the last century several experiments
 were performed to test the predictions of the Thermo-acoustic Model using
 accelerator- or laser beams \cite{SUL,GOL}. 
The basic experimental arrangement is shown in fig.~\ref{fig:sulak}. 

An intense laser or low energy proton beam is stopped in a water tank. 
Varying the beam parameters allowed several model checks like the following:
\begin{itemize}
\setlength{\itemsep}{0.5pt}
\item  varying intensity \hspace{0.5cm} ($E$-dependence)    
\item  varying beam diameter \hspace{0.5cm} ($f$-dependence)
\item  varying distance \hspace{0.5cm} ($R$-dependence)
\item  varying liquids \hspace{0.5cm} ($\gamma$-dependence)
\end{itemize}

During the last decade several similar tests have been made by different
 groups for water \cite{ITEP, ECAP-06, ROM}, ice \cite{BOS-03, ST-04} and
 permafrost \cite{NRT}. 
The general conclusions from all these tests are the following:
\begin{itemize}
\setlength{\itemsep}{0.5pt}
\item many predictions of the Thermo-acoustic Model could be confirmed
\item the dominant mechanism for acoustic signal production is thermal expansion
\item other contributions, e.g. from micro-bubble formation could not
 completely be excluded
\item absolute signal values could not be checked, because  the energy deposit
 in the different tests was difficult to calculate precisely and was different
 to the ultra-high energetic particle case.
\end{itemize}

\begin{figure}
  \centering
  \includegraphics[width=8.0cm]{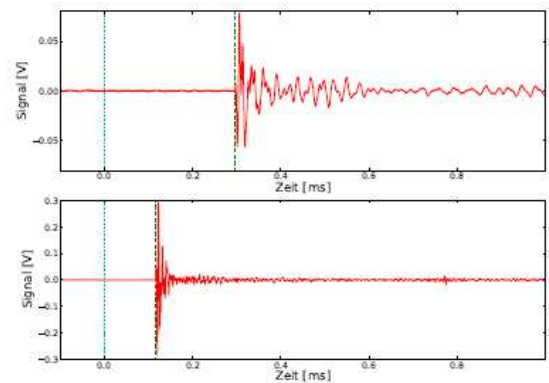}
  \caption{\label{fig:dirk} Acoustic signals from a laser beam shot in a tank
 filled with water (top) and ice (bottom)  (from \cite{AAL-1})}
\end{figure}

An unsolved problem was until recently also the relative signal strength in
 different target materials. 
It is therefore very valuable that an experiment is performed this year at the
 Aachen-Acoustic-Laboratory aiming to answer this question \cite{AAL-1}. 
A well controlled laser beam is stopped in a large tank placed in an even
 larger freezer. 
This allows measurements in a temperature range from +20 to -25 degree Celsius
 in water and ice. 
In fig.~\ref{fig:dirk} acoustic signals for both configurations are shown. 
The time difference of the signals is due to the different velocity of sound in
 both materials. 
The signal strength in ice is about a factor six larger than in water
 consistent with the theoretical expectation. The exact comparison has
 however to be done for the corresponding pressure amplitudes.
More detailed results of these measurements will be available soon.

\section{Acoustic transducers}

Acoustic transducers for ultra-sound have been developed since long for
 applications in water. 
Their quality profited from their use in military projects. 
Today commercial products are available from several companies. 
Special requirements like stability at high pressure for use at large water
 depth lead to considerably high prices per piece. 
Because large numbers of corresponding devices are necessary to build large
 detector arrays, several attempts were started to build own sensitive sensors
 and transmitters. 
For applications in ice, salt and permafrost this was unavoidable anyway. 

 \subsection{piezo ceramic sensors}

\begin{figure}
  \centering
  \includegraphics[width=8.0cm]{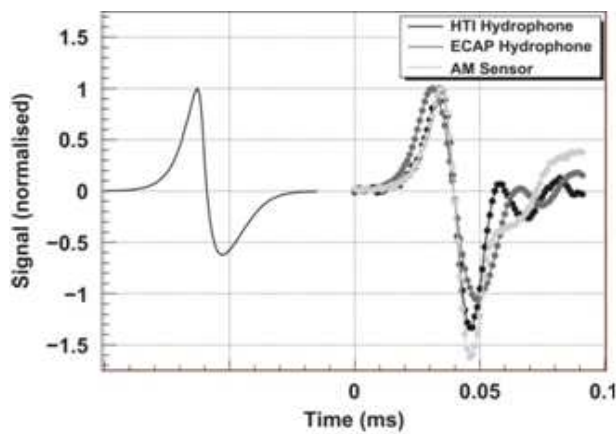}
  \caption{\label{fig:ecap-10} Comparison of the response of three different acoustic sensor types to a bipolar pulse (from \cite{ECAP-10})}
\end{figure}

Nearly all acoustic sensors in use in todays test arrays are based on
 piezo-ceramic elements \cite{ECAP-06-2}. 
Within the AMADEUS project a nice comparison of different commercial and
 self-made devices has been performed \cite{ECAP-10}. 
In fig.~\ref{fig:ecap-10} the response of three of their sensors to the same
 transmitter signal is shown. An interesting development is their
 ``Acoustic Module'', where piezo-ceramic elements are integrated in a glass
 pressure sphere otherwise used to contain photomultipliers for light
 detection in the ANTARES experiment \cite{ANTA}.  

\begin{figure}
  \centering
  \includegraphics[width=8.0cm]{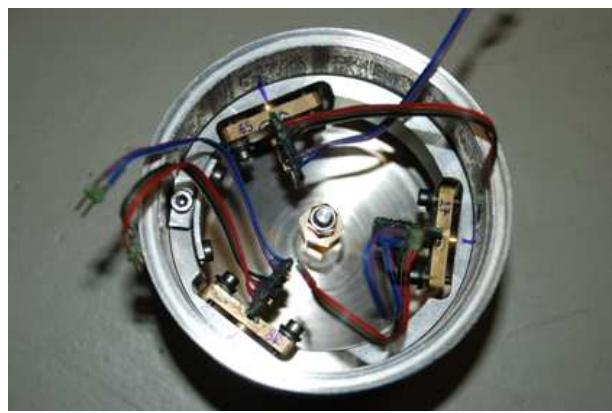}
  \caption{\label{fig:spatsd} SPATS second generation sensor module  (from \cite{SPATS-T})}

\end{figure}

The SPATS group connected to the IceCube experiment started their own
 developments for ice applications with a similar concept \cite{BOS-03,BOS-05}
 but changed their design finally to a cylindrical steel pressure housing where
 three piezo-ceramics are pressed against the inner wall
 (see fig.~\ref{fig:spatsd}). 
Typical sensitivities of sensors in use are in the range -190 to
 -110 dB re 1V/$\mu$Pa.

Recently the use of piezo-foils for the design of acoustic sensors has been
 reported \cite{AAL-1}. 
This would have the advantage that one would be more flexible in sensor shape
 and size. 
Results from forthcoming tests will be hopefully published soon.

\subsection{other sensor concepts}

A ``new'' concept for acoustic sensors under discussion in  particle detection
 applications since a few years is the use of acousto-optic hydrophones. 
In fact the corresponding working principles using either frequency or
 intensity modulations in optical fibers has been discussed already more
 than 30 years ago \cite{FIELD}.  

Recently several groups published results for a fiber laser, where the
 modulation of the cavity size by pressure pulses leads to modulations of the
 laser wavelength detectable by interferometry \cite{FIBLAS}. 
The concept has been proven to work at a static pressure of 35 MPa, i.e. 3500 m
 depth with higher sensitivity and better resolution than piezo-ceramic based
 devises. 
An application in a real open water test array is however still missing.

The application of coupled waveguide intensity modulated hydrophones is under
 discussion for an upgrade of the BAIKAL acoustic test setup \cite{BUD}.

 \subsection{transmitters}

Acoustic transmitters are normally used in acoustic arrays to get position and
 calibration information. 
An exceptional idea was realized for this purpose in the SAUND-1 experiment,
 where light bulbes of different size were deployed on top of seven hydrophones
 at the AUTEC military array near the Bahamas \cite{JUST}.

The SPATS group published results from different types of pingers fixed in ice
 or used in water filled holes to measure the acoustic attenuation length in
 South Pole ice \cite{SPATS-A}. 
A byproduct of these measurements was the observation of shear waves in the
 ice \cite{SPATS-V}.

Position monitoring is also a problem of all deep water optical neutrino
 telescopes. 
It is normally solved by fixing strong acoustic transmitters at the sea bed and
 by adding some hydrophones to the optical strings. 
Pulses emitted with short time differences allow then precise position
 monitoring at the 10 cm scale \cite{ECAP-POS}. 
For the planned KM3Net detector a corresponding system is under development,
 which will allow to observe  pulses from particle cascades with high
 sensitivity \cite{RICO-10}. 

To test the sensitivity of acoustic arrays in-situ is still an unsolved
 problem. 
Two groups try to develop acoustic pulser systems which mimic pulses from
 ultra-high energetic neutrino interaction in strength and shape. 
Successful tests have been made using arrays with more than 5 transmitters
 taking into account the transfer function between emitted and received signals
 \cite{DAN-10}. 
In another approach a parametric acoustic source is used. 
The overlay of two high frequency signals fed to an emitter leads to the
 production of a bipolar pulse at lower frequency whereas the remaining high
 frequency components are quickly absorbed in the medium \cite{BOUCAB}.


\section{Sensor calibration}

For the application of the acoustic technology for particle detection the use
 of carefully calibrated sensors is mandatory. 
Several methods are applied for this purpose in the laboratory \cite{ARD-08},
 e.g. comparison with a calibrated reference hydrophone and reciprocity
 calibration or use of calibrated emitters. 
With the last method all sensors used for the AMADEUS project were calibrated
 in dependence of the azimuthal and zenith angle in a water tank
 \cite{ECAP-10}. 
One has to keep in mind, however, that the sensor sensitivity depends on the
 specific environmental conditions at the deployment location as e.g.
 temperature and pressure. 
Together with a NATO institute the ONDE-group has developed a standard
 procedure for under pressure calibration. 
They reported a sensitivity change of about 2 dB/1000 m water depth
 \cite{RICO-11}.  

In-situ calibration in ice is even more complicated than in water. 
Beside pressure effects, deep temperature and impedance changes between ice and
 water have to be taken into account. 
The SPATS group has calibrated their sensors in water and studied separately
 the dependence on pressure (in a water-oil mixture), temperature (in air) and
 impedance (measuring noise level changes during sensor freeze-in). 
Assuming that the corresponding sensitivity changes can just be multiplied, 
 they got a final result with about 40 percent error \cite{SPATS-N}.

\begin{figure}
  \centering
  \includegraphics[width=8.0cm]{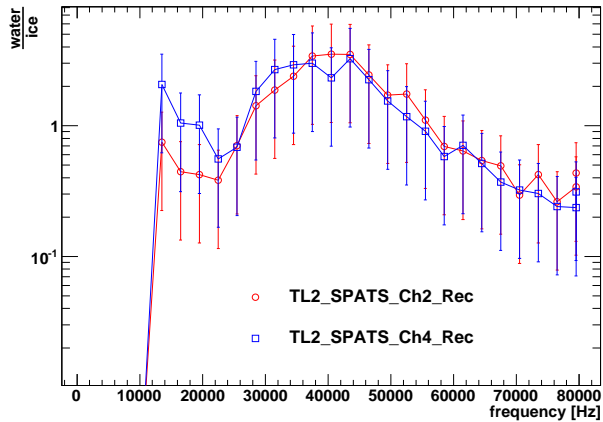}
  \caption{\label{fig:VergleichWasserEis} Preliminary result for the ratio of
 sensitivities of a third generation SPATS sensor deployed in water and ice
 (from \cite{AAL-3})}
\end{figure}

A direct comparison of calibration results in water and ice at normal pressure
 was done for a third generation SPATS sensor in the Aachen-Acoustic Laboratory
 \cite{AAL-3}. 
Preliminary results of a reciprocity calibration, shown in
 fig.~\ref{fig:VergleichWasserEis} are consistent with the estimation described above. 
The final result of these measurements is expected to be published soon.

\section{Acoustic test arrays}

During the last decade results were reported from six test sites either using
 part of military hydrophone arrays or installing own acoustic sensors. 
The main purpose of these studies was to clarify how sensitive acoustic
 detectors could be to signals from neutrino interactions at given
 environmental conditions like acoustic noise, signal attenuation and
 refraction. 
A few prominent results of the different groups will be discussed below. 
It is however out of the scope of this presentation to give a detailed overview
 about presently available data. 
To get more information the reader is pointed to the references given in this
 paper and to the proceedings of the ARENA conferences since 2005 \cite{ARENAS}.

 \subsection{SAUND}

The ``Study of Acoustic Ultra-high Neutrino Detection - SAUND'' started in
 2003. 
Seven hydrophones of the military AUTEC array near the Bahamas were used to
 search for acoustic signals. In 196 days $65 \times 10^6$ triggers were taken
 in a sensitive volume of about 15 ${\rm km^3}$. 
The data were used to calculate the first acoustic neutrino flux limit \cite{JUST}.

\begin{figure}
  \centering
  \includegraphics[width=8.0cm]{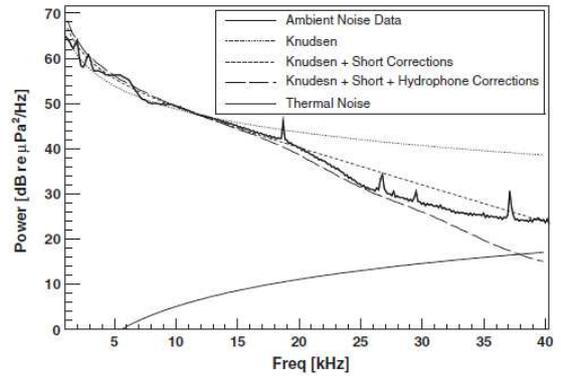}
  \caption{\label{fig:KURA-noise} Average ambient noise level measured by SAUND
  (from \cite{KURA-1})}
\end{figure}

In the second phase of this experiment the number of used hydrophones increased
 to 49 and the volume under study to 1500 ${\rm km^3}$. 
130 days of data taking allowed detailed studies of acoustic noise behaviour
 (see fig.~\ref{fig:KURA-noise}) \cite{KURA-1}.  
In a complex data reduction and signal processing procedure two events were
 found compatible with showers from neutrino interactions above $10^{22}$ eV. 
The result was used to derive an improved neutrino flux limit
 (see fig.~\ref{fig:Nu-flux}) \cite{KURA-2}.

\subsection{Acorne}

The ``Acoustic COsmic Ray Neutrino Experiment - ACORNE'' is an activity of
 different groups in the UK. 
Eight hydrophones of the RONA military array in North-West Scotland are used
 for acoustic signal searches. 
Between 2006 and 2008 nearly 30 Tb of raw data have been collected. 
Advanced signal processing and filtering technologies were developed by the
 group and applied to the data \cite{DAN-06,BEV-08}. 
Corresponding data reduction and efficiency calculations allowed to derive a
 neutrino flux limit in the same energy range as the SAUND result
 (see fig.~\ref{fig:Nu-flux}). 
Furthermore information about signal attenuation and localization was
 collected. 
The examination of the data for black hole signatures lead to the calculation
 of a corresponding upper limit \cite{DAN-10-1}.

\subsection{ONDE}

Within the ``Ocean Noise Detection Experiment - ONDE'' four hydrophones
 forming a tetrahedron were deployed in 2005 to about 2000 m depth in the
 Mediterranean See 25 km offshore of Catania/Italy.
Noise was monitored about five minutes every hour. 
The average noise level in the 20-43 kHz frequency region was found to be
 5.4 $\pm$ 2.2 $\pm$ 0.3 mPa \cite{RICO-09}. 
The noise level was strongly correlated with the actual environmental
 conditions at the sea surface. 

\begin{figure}
  \centering
  \includegraphics[width=8.0cm]{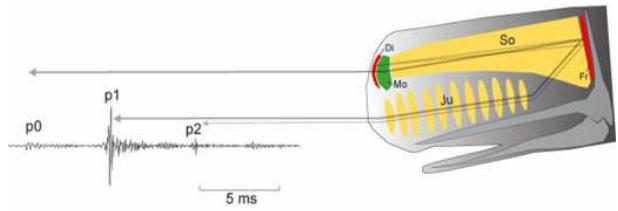}
  \caption{\label{fig:spermwhales} Acoustic signals from a sperm whale  (from \cite{RICO-11})}

\end{figure}

Signals observed in the noise data more often than expected were produced by a
 population of sperm whales. 
Analyzing these signal in detail (for an example see fig.
 ~\ref{fig:spermwhales}) even allowed to derive age and gender of the animal
 emitting it \cite{RICO-11}. 
Corresponding studies will continue in collaboration with marine biologists.

\subsection{Baikal}

\begin{figure}
  \centering
  \includegraphics[width=8.0cm]{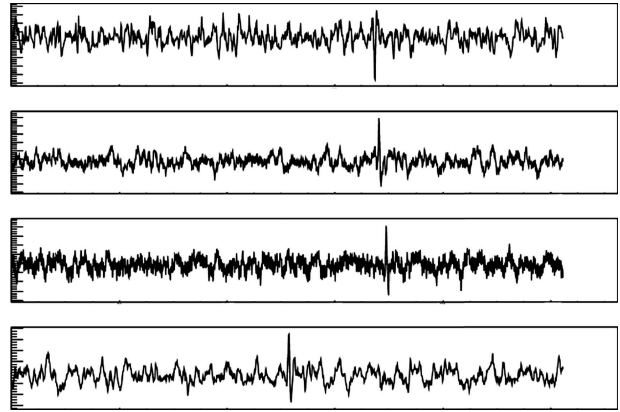}
  \caption{\label{fig:Baikal-event} Acoustic event reconstructed as upward going neutrino-like signal  (from \cite{BUD})}
\end{figure}

A subgroup of the collaboration which constructed and operates the first
 optical neutrino telescope in lake Baikal \cite{BAIKAL} has deployed a
 digital hydro-acoustic module with four hydrophones at a regular tetrahedron
 of 1.5 m edge length in 150 m depth at one of the outer strings of the
 optical array. 
With this device and its predecessors extensive environmental studies have
 been done. 
The noise level was measured most of the time below 5 mPa. 
Accepting only signals from the deep lake one interesting neutrino-like event
 has been observed (see fig. ~\ref{fig:Baikal-event}).

In March 2011 an acoustic string with three acoustic modules has been deployed
 \cite{BUD}. 
Data taking is ongoing and first results are expected to be shown at next years
 conferences. 

 \subsection{ANTARES}

A detailed description of the design and performance of the
 ``ANTARES Module for the Acoustic DEtection Under the Sea - AMADEUS'' can be
 found in \cite{ECAP-10}. 
36 sensors are located at three storeys of both string 12 and the
 instrumentation line of the ANTARES optical neutrino telescope \cite{ANTA}
 located at a depth of $\sim 2500$ m in the Mediterranean sea about 40 km
 distant to Toulon/France. 
Data taking started in 2007. 
Extensive noise studies have shown strong correlations to actual weather
 conditions. 
A high statistics frequency dependent noise measurement reflects similar
 results like those found with ONDE in the deep sea near Sicily (see
 section 6.3).

\begin{figure}
  \centering
  \includegraphics[width=8.0cm]{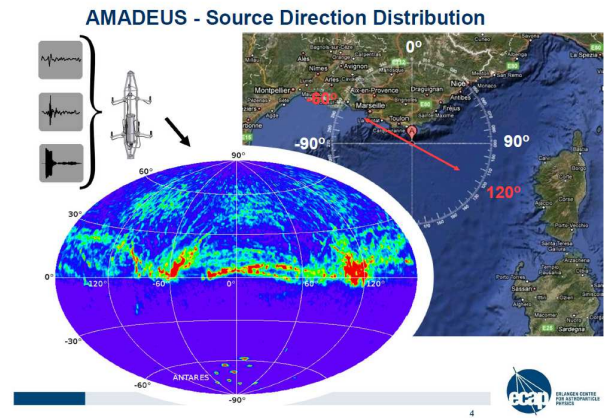}
  \caption{\label{fig:ANTA-TRANS} Angular distribution of transient events observed with the AMADEUS detector(from \cite{LAH-11})}
\end{figure}

The arrival direction of transient acoustic signals could be determined (see
 fig. ~\ref{fig:ANTA-TRANS}) \cite{LAH-10}. 
It was found that these signals have mostly antroprogenic origin being due to
 heavy ship traffic in this region of the sea.

\subsection{SPATS}

The ``South Pole Acoustic Test Setup - SPATS'' has been deployed in the upper
 empty part of holes drilled for the IceCube neutrino observatory
 \cite{IceCube-1}. 
Seven acoustic stations with transmitters and receivers are positioned between
 80 m and 500 m in the ice at the South Pole, with a maximum distance of about
 520 m between strings \cite{SPATS-T}. 
Data taking started in early 2007. 
Since then results have been published for the speed of sound of pressure and
 shear waves and their refraction versus depth \cite{SPATS-V}, the acoustic 
 attenuation length \cite{SPATS-A} and the ambient noise level \cite{SPATS-N}.

\begin{figure}[hbt]
  \centering
  \includegraphics[width=8.0cm]{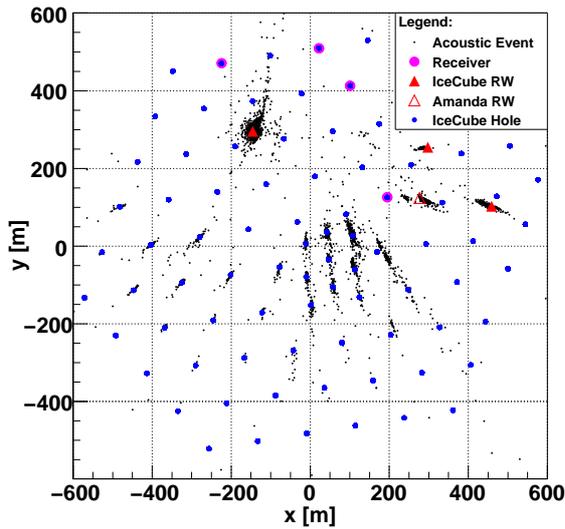}
  \caption{\label{fig:all-4s-XY-con} Distribution of transient signal source locations in the x-y plane compared with the position of IceCube holes and Rod-wells (from \cite{SPATS-N})}
\end{figure}
 
Transient events were observed from re-freezing IceCube holes and from
 water reservoirs used for hole drilling
 (see fig.~\ref{fig:all-4s-XY-con}). 
From the non-observation of acoustic signals in a region outside the IceCube
 construction area a neutrino flux limit has been estimated
 (see fig.~\ref{fig:Nu-flux}).


\section{Acoustic neutrino-flux limits}

\begin{figure}
  \centering
  \includegraphics[width=8.0cm]{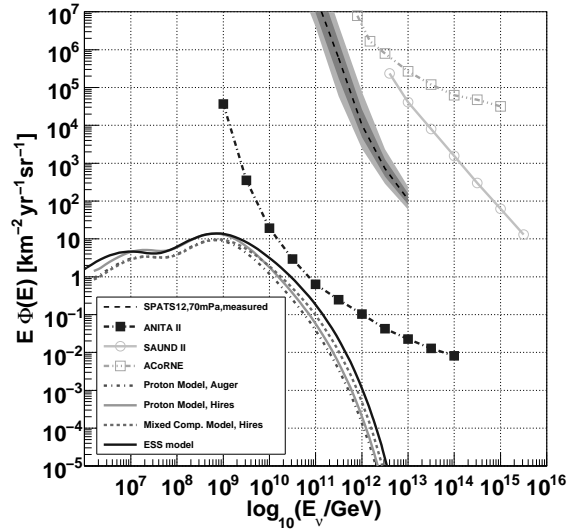}
  \caption{\label{fig:Nu-flux} Available neutrino flux limits from acoustic test arrays compared to some model predictions and results from some radio detectors sensitive at highest energies (from \cite{SPATS-N})}
\end{figure}

Neutrino flux limits have been derived until now in three acoustic test
 experiments \cite{KURA-2,BEV-08,SPATS-N}, all not optimized or not designed
 for this purpose. 
In fig. ~\ref{fig:Nu-flux} the corresponding results are shown and compared
 with predictions of one frequently quoted cosmogenic neutrino flux model and
 its recent modifications \cite{ESS}. 
The presently best limit in the considered energy region has been published
 recently by the ANITA collaboration using radio antennas as payload of a
 balloon circling in a hight of ~35 km around Antarctica \cite{ANITA}. 
The detector was searching for pulses of coherent radio emission from neutrino
 induced cascades in the ice below in a volume of about 1.6$\times 10^6$
 ${\rm km^3}$.

The acoustic neutrino flux limits are today still more than four orders of
 magnitude less sensitive than the best radio limits which is partly explained
 by the huge difference in the corresponding detection volumes used. 
What makes all limits difficult to interpret are the partly unknown systematic
 errors of the measurements and the reliability of assumptions made for
 efficiency and sensitivity calculations. 
This has been discussed e.g.in \cite{SPATS-N}.

\section{Future activities}  

The experiments mentioned in section 6 have in the meantime either finished or
 achieved their primary goals. 
Follow up programs are under discussion or already in a planning stage.

\subsection{acoustic detection in water}

With the present facilities ``fake neutrino sources'' (see section 4.3) will be
 used to get better knowledge about in-situ detector efficiencies and trigger
 schemes. 
On an intermediate time scale design studies towards a ${\rm km}^3$-sized
 hybrid acousto-optic detector will hopefully be followed by its construction
 in the Mediterranean KM3Net project \cite{RICO-11}. 
Most European acoustic activities should converge around this program.

\subsection{acoustic detection in ice}

Ice provides probably the best conditions to build a hybrid
 radio-acoustic-optical detector allowing background-free detection of the
 rare neutrino signals above $10^{18}$ eV. 
Having evaluated the acoustic ice properties at the South Pole, the boundary
 conditions for such a detector are now mostly known at that location. 
Corresponding Monte Carlo design studies are under way. 
They show, that already a 100 ${\rm km}^3$ scale detector would need a large
 number of holes to be drilled. 
The development of robotic drilling and deployment methods is mandatory for
 such a project. 
First attempts in that direction are just starting \cite{LAIH-11}.


The detection of ultra-high energy neutrinos and the detailed study of their
 interactions remains a top priority topic of astroparticle physics also in the
 long term future. 
Corresponding experiments need most probably to employ multiple detection
 methods with different systematics for successful operation. 
The acoustic particle detection technology is qualified to be one of them.

\section*{Acknowledgements} 

I would like to thank the organizers of this conference for the invitation to
 give this presentation. 
I thank also N. Budnev, G. Gratta, D. Heinen, R. Lahmann and G. Riccobene for
 providing new information about activities of their groups. I am indebted to J. Berdermann for valuable suggestions and critical reading of the manuscript.








\end{document}